\documentstyle[11pt]{article}
\advance \topmargin by -\headheight
\advance \topmargin by -\headsep
     
\evensidemargin \oddsidemargin
\def\rf#1{(\ref{eq:#1})}
\def\lab#1{\label{eq:#1}}

\def\br{\begin{eqnarray}}
\def\er{\end{eqnarray}}
\def\be{\begin{equation}}
\def\ee{\end{equation}}
\def\eq{\!\!\!\! &=& \!\!\!\! }
\def\foot#1{\footnotemark\footnotetext{#1}}

\def\llb{\left\lbrack}
\def\rrb{\right\rbrack}

\def\lcurl{\left\{}
\def\rcurl{\right\}}
\def\({\left(}
\def\){\right)}

\def\lskip{\vskip\baselineskip\vskip-\parskip\noindent}
\newcommand\partder[2]{{{\partial {#1}}\over{\partial {#2}}}}

\newcommand\sbr[2]{\left\lbrack\,{#1}\, ,\,{#2}\,\right\rbrack}
\newcommand\Sbr[2]{\Bigl\lbrack\,{#1}\, ,\,{#2}\,\Bigr\rbrack}

\def\a{\alpha}
\def\b{\beta}

\def\d{\delta}

\def\l{\lambda}

\def\m{\mu}

\def\o{\over}

\def\P{\Phi}
\def\pa{\partial}

\def\bpa{{\bar \partial}}
\def\pr{\prime}

\def\t{\tau}

\newcommand\sumi[1]{\sum_{#1}^{\infty}}

\def\cF{{\cal F}}

\def\cL{{\cal L}}
\def\cM{{\cal M}}

\def\cW{{\cal W}}

\def\mark{\noindent{\bf Remark.}\quad}

\newtheorem{proposition}{Proposition}

\newtheorem{lemma}{Lemma}
\newtheorem{corollary}{Corollary}
\def\proof{\par{\it Proof}. \ignorespaces} \def\endproof{{$\Box$}\par}

\newcommand\DB{{Darboux-B\"{a}cklund}~}

\def\Res{{\rm Res}}

\def\vp{{\varphi}}
\newcommand\BA{\psi_{BA} (t,\l)}   
\newcommand\BAc{\psi_{BA}^{\ast} (t,\l)}

\def\bt{{\bar t}}
\def\pai{\partial^{-1}}
\def\bD{{\bar D}}
\def\bpai{{\bar \partial}^{-1}}
\def\bcL{{\bar {\cal L}}}
\def\bP{{\bar \Phi}}
\def\bPsi{{\bar \Psi}}

\newcommand{\ct}[1]{\cite{#1}}
\newcommand{\bi}[1]{\bibitem{#1}}

\newcommand\NPB[3]{{\sl Nucl. Phys.} {\bf B#1} (#2) #3}

\newcommand\CMP[3]{{\sl Commun. Math. Phys.} {\bf #1} (#2) #3}

\newcommand\IJMPA[3]{{\sl Int. J. Mod. Phys.} {\bf A#1} (#2) #3}

\newcommand\PHSA[3]{{\sl Physica} {\bf A#1} (#2) #3}

\begin{document}

\vspace*{-1.5cm}
\noindent
{\sl solv-int/9712012} \hfill{BGU-97/21/Dec-PH}\\
\phantom{bla}
\hfill{UICHEP-TH/97-16}\\
\begin{center}
{\large {\bf A New ``Dual'' Symmetry Structure of the KP Hierarchy}}
\end{center}
\vskip .15in
\begin{center}
{ H. Aratyn\footnotemark
\footnotetext{Work supported in part by the U.S. Department of Energy
under contract DE-FG02-84ER40173}}
\par \vskip .1in \noindent
Department of Physics, University of Illinois at Chicago\\
845 W. Taylor St., Chicago, IL 60607-7059, U.S.A.\\
\par \vskip .15in
{ E. Nissimov$^{2}$  and
S. Pacheva\foot{Supported in part by Bulgarian 
NSF grant {\em Ph-401}}}
\par \vskip .1in \noindent
Institute of Nuclear Research and Nuclear Energy \\
Boul. Tsarigradsko Chausee 72, BG-1784 $\;$Sofia, Bulgaria \\
\vspace{-0.1in}
\begin{center} and \end{center}
\vspace{-0.1in}
Department of Physics, Ben-Gurion University of the Negev \\
Box 653, IL-84105 $\;$Beer Sheva, Israel \\
\end{center}

\begin{abstract}
A new infinite set of commuting additional (``ghost'') symmetries
is proposed for the KP-type integrable hierarchy.
These symmetries allow for a Lax representation in which they 
are realized as standard isospectral flows.
This gives rise to a new double-KP hierarchy embedding 
``ghost'' and original KP-type Lax hierarchies  connected to each other
via a ``duality'' mapping 
exchanging the isospectral and ``ghost'' ``times''.
A new representation of 2D Toda lattice hierarchy as a special \DB
orbit of the double-KP hierarchy is found and parametrized
entirely in terms of (adjoint) eigenfunctions of the original KP subsystem.
\end{abstract}
\lskip
We first provide some background information on the KP hierarchy and ``ghost'' 
symmetries.
In what follows we use the Sato formalism of pseudo-differential operator
calculus (see, e.g. \ct{ldickey}) to describe Kadomtsev-Petviashvili (KP) 
type integrable hierarchies
of integrable nonlinear 
evolution equations. The main object is
the pseudo-differential Lax operator $\cL$ obeying an infinite set of
evolution equations \foot{
We shall employ the following notations:
for any (pseudo-)\-differential operator $A$ and a function $f$, the symbol
$\, A(f)\,$ will indicate 
action of $A$ on $f$, whereas the
symbol $Af$ will denote just operator product of $A$ with the zero-order
(multiplication) operator $f$. The symbol $D$ stands for the differential
operator $\pa/\pa x$, whereas $\pa \equiv \pa_x$ will denote derivative on a
function. Further, in what follows
the subscripts $(\pm )$ of any pseudo-differential
operator $A = \sum_j a_j D^j$ denote its purely differential part
($A_{+} = \sum_{j\geq 0} a_j D^j$) or its purely pseudo-differential part
($A_{-} = \sum_{j \geq 1} a_{-j} D^{-j}$), respectively.}
w.r.t. the KP ``times'' 
$(t) \equiv (t_1 \equiv x, t_2 ,\ldots )$ :
\be
\cL = D + \sum_{i=1}^{\infty} u_i D^{-i} \qquad ; \qquad
\partder{\cL}{t_l} = \Sbr{\(\cL^{l}\)_{+}}{L} \quad , \; \;
l = 1, 2, \ldots
\lab{lax-eq}
\ee
Equivalently, one can represent \rf{lax-eq} in terms of the 
dressing operator $W$ whose pseudo-differential series are directly 
expressed in terms of the so called tau-function $\t (t)$ :
\be
\cL = W D W^{-1} \quad ,\quad \partder{W}{t_l} = - \(\cL^l\)_{-} W \quad ,
\quad W = \sum_{n=0}^{\infty} \frac{p_l \( - [\pa]\)\t (t)}{\t (t)} D^{-l}
\lab{W-main}
\ee
with the notation: $[y] \equiv \( y_1,  y_2/2 , y_3/3 ,\ldots \)$
for any 
multi-variable $(y) \equiv \( y_1 ,y_2 ,y_3 ,\ldots \)$ and with $p_k (t)$
being the Schur polynomials. The tau-function is related to the Lax operator
as:
\be
\pa_x \partder{}{t_l}\ln \t (t) = {\rm Res} \cL^{l} \; .
\lab{tau-L}
\ee
In the present approach a crucial notion is that of (adjoint) eigenfunctions 
((adj-)EFs) $\P (t),\, \Psi (t)$ of the KP hierarchy satisfying:
\be
\partder{\Phi}{t_k} = \cL^{k}_{+}\bigl( \Phi\bigr) \qquad; \qquad
\partder{\Psi}{t_k} = - \(\cL^{*} \)^{k}_{+}\bigl( \Psi\bigr)
\lab{eigenlax}
\ee
The (adjoint) Baker-Akhiezer (BA) ``wave''
functions $\psi_{BA} (t,\l ) = W (\exp(\xi (t,\l )))$
and $\psi_{BA}^{*} (t,\l )= (W^{*})^{-1}(\exp(-\xi (t,\l )))$ ( with
$\xi (t,\l ) \equiv \sum_{l=1}^\infty t_l \l^l$)
are (adj-)EFs which, in addition, also satisfy the spectral equations
of the form
$\cL^{(*)}\bigl( \psi_{BA}^{(*)} (t,\l )\bigr) =  \l \psi_{BA}^{(*)} (t,\l )$.
As shown in \ct{ridge}, any (adj-)EF possesses a spectral representation of
the form:
\be
\P (t) = \int d\l \, \vp (\l )\, \BA \quad , \quad
\Psi (t) = \int d\l \, \vp^{*} (\l )\, \BAc
\lab{spec}
\ee
with certain suitable spectral ``densities'' $\vp^{(*)} (\l )$.

The so called {\em squared eigenfunction
potential} $S(\P ,\Psi )$ \ct{oevela} yields a well-defined unique  
expression for the inverse derivative $\pa_x^{-1}$ of a
product of arbitrary pair of EF and adj-EF \ct{ridge} :
\be
\pai (\P (t) \Psi (t)) \equiv S (\P ,\Psi ) = 
- \int\int\, d\l\, d\m \, \vp^{\ast}(\l ) \, \vp (\m ) 
{e^{\xi \( t, \m \)- \xi(t,\l)} \o \l -\m}\, { e^{\sumi{1} {1 \o l}
\( \l^{-l} - \m^{-l} \) \partder{}{t_l} } \t (t) \o \t (t)}
\lab{S-def} 
\ee
This will always be the case for all instances of appearance of inverse
derivatives in the sequel.

Finally, let us recall the basic facts about ``ghost'' symmetries of 
the generic KP hierarchy. 
A ``ghost'' symmetry is defined through an action of a vector field
${\hat \pa}_\a$ on the KP Lax operator or the dressing operator \ct{orlov} :
\be
{\hat \pa}_\a \cL = \Sbr{\cM_\a}{\cL} \quad , \quad
{\hat \pa}_\a W = \cM_\a W \quad , \quad
\cM_\a \equiv \sum_{a \in \{\a\}} \P_a D^{-1} \Psi_a
\lab{ghost-flow-Lax}
\ee
where $\(\P_a ,\Psi_a\)_{a \in \{\a\}}$ are some set of functions
indexed by $\{\a\}$. Commutativity of
${\hat \pa}_\a $ with $\partder{}{t_l}$ 
implies that
$\(\P_a ,\Psi_a\)_{a \in \{\a\}}$ is a set of pairs of (adj-)EFs of $\cL$.

Now, for the general (adj-)EFs $\P ,\Psi$ of $\cL$ we define new generalized
``ghost'' symmetry flows : 
\be
{\hat \pa}_\a \P = \sum_{a \in \{\a\}} \P_a \pai \(\Psi_a \P\) - \cF^{(\a )}
\quad ,\quad
{\hat \pa}_\a \Psi = \sum_{a \in \{\a\}} \Psi_a \pai \(\P_a \Psi\) +
{\cF^{\ast}}^{(\a )} \, .
\lab{ghost-flow-eigenf}
\ee
Note the additional inhomogeneous terms 
$\cF^{(\a )},\, {\cF^{\ast}}^{(\a )}$
which themselves are (adj-)EFs of $\cL$ \rf{lax-eq} and which 
are absent in the traditional approach of \ct{orlov,oevela},
see however \ct{EOR95}. 
It is crucial for what follows that their presence is in general
allowed by requirements of 
{\em commutativity of two different ``ghost'' flows} ${\hat \pa}_\a$ and
${\hat \pa}_\b$ and the integrability condition
$\sbr{{\hat \pa}_\a - \cM_\a}{{\hat \pa}_\b - \cM_\b}=0$ 
following from the definition \rf{ghost-flow-Lax}. 

We now proceed to give an explicit construction of 
the ``ghost'' KP hierarchy.
Consider an infinite system of independent (adj-)EFs
$\lcurl \P_j ,\Psi_j\rcurl_{j=1}^{\infty}$ of $\cL$ and define
the following infinite set of the ``ghost'' symmetry flows:
\br
\partder{}{\bt_s} \cL &=& \Sbr{\cM_s}{\cL} \qquad ,\qquad
\cM_s = \sum_{j=1}^s \P_{s-j+1} D^{-1} \Psi_j 
\lab{ghost-s}  \\
\partder{}{\bt_s} \P_k \eq \sum_{j=1}^s \P_{s-j+1} \pai \(\Psi_j \P_k\) \; -\;
\P_{k+s} \quad ; \quad
\partder{}{\bt_s} \Psi_k = \sum_{j=1}^s \Psi_j \pai \( \P_{s-j+1} \Psi_k\)
\; + \; \Psi_{k+s} 
\lab{M-s-eigenf} \\
\partder{}{\bt_s} F &=& \sum_{j=1}^s \P_{s-j+1} \pai \(\Psi_j F\) \quad ; \quad
\partder{}{\bt_s} F^\ast = \sum_{j=1}^s \Psi_j \pai \( \P_{s-j+1} F^\ast\)
\lab{M-s-generic}
\er
where $s,k =1,2,\ldots$ , and ($F^{*}$) 
$F$ denote generic (adj-)EFs which do not belong to the ``ghost''
symmetry generating set $\lcurl \P_j ,\Psi_j\rcurl_{j=1}^{\infty}$. 
With the choice of the inhomogeneous terms as in
\rf{M-s-eigenf} it is easy to show that the `ghost'' symmetry flows
$\partder{}{\bt_s}$ do indeed commute, {\sl i.e.}, the
$\pa$-pseudo-differential operators $\cM_s $ \rf{ghost-s} satisfy:
\be
\partder{}{\bt_s} \cM_r - \partder{}{\bt_r} \cM_s - \Sbr{\cM_s}{\cM_r} = 0 \; .
\lab{ZS-cM-s}
\ee
In particular, for the first ``ghost'' symmetry flow 
$\partder{}{\bt_1} \equiv \bpa$ we have:
\be
\bpa \P_k = \P_1 \pai \(\Psi_1 \P_k\) - \P_{k+1} \quad , \quad
\bpa \Psi_k = \Psi_1 \pai \(\P_1 \Psi_k\) + \Psi_{k+1} \quad ; \quad
\bpa F = \P_1 \pai \(\Psi_1 F\) 
\lab{ghost-1}
\ee
Eqs.\rf{ghost-1}, in turn, imply the following 2D Toda lattice 
(2DTL)-like equations for the 
Wronskians of $\P_j$'s and $\Psi_j$'s, respectively:
\be
\pa\bpa\ln W_k = \P_1 \Psi_1 - \frac{W_{k+1} W_{k-1}}{{W_k}^2}
\quad ; \quad
\pa\bpa\ln \cW_k = \P_1 \Psi_1 + \frac{\cW_{k+1} \cW_{k-1}}{{\cW_k}^2} \; .
\lab{2DTL-like} 
\ee
Here and below use will be made of the following short-hand notations for
the Wronskian-type determinants:
\br
W_k \equiv W_k \llb \P_1 ,\ldots ,\P_k \rrb =
\det {\Bigl\Vert} \pa^{\a -1} \P_\b {\Bigr\Vert} \;\;\;, \;\;
\a ,\b =1,\ldots ,k \quad ; \quad
\cW_k \equiv W_k \llb \Psi_1 ,\ldots ,\Psi_k \rrb
\lab{W-notation} \\
W_k (F) \equiv W_{k+1} \llb \P_1 ,\ldots ,\P_k ,F \rrb \quad , \quad
\cW_k (F^\ast ) \equiv W_{k+1} \llb \Psi_1 ,\ldots ,\Psi_k ,F^\ast \rrb
\lab{W-notation-1}
\er

Consider now the $\t$-function of $\cL$ 
and let us act with $\partder{}{\bt_s}$ on both sides of \rf{tau-L} obtaining:
\be
\partder{}{\bt_s} \ln \t = - \sum_{j=1}^s \pai \(\P_{s-j+1} \Psi_j \)
\lab{tau-bt-s}
\ee
using \rf{ghost-s} as well as the $t_r$-flow eqs.
$\partder{}{t_r}\cM_s = {\Sbr{\cL^r_{+}}{\cM_s}}_{-}$.
In particular, for $s=1$ eq.\rf{tau-bt-s} together with \rf{ghost-1} yields:
\be
\bpa \ln \t = - \pai \(\P_1 \Psi_1 \) \quad ,\quad
\bpa \ln \(\P_1 \t\) = - \frac{\P_2}{\P_1} \quad , \quad
\bpa \ln \(\Psi_1 \t\) = \frac{\Psi_2}{\Psi_1}
\lab{tau-P-bt-1}
\ee
Taking into account the first eq.\rf{tau-P-bt-1}, we can rewrite 
\rf{2DTL-like} in the standard 2DTL form: 
\be
\pa\bpa\ln \( W_k \t\) =  - \frac{\( W_{k+1}\t\)\( W_{k-1}\t\)}{\( W_k \t\)^2}
\quad ; \quad
\pa\bpa\ln \(\cW_k \t\) =
\frac{\(\cW_{k+1} \t\)\(\cW_{k-1} \t\)}{\(\cW_k \t\)^2} \; .
\lab{2DTL-1}
\ee

Using last of eqs.\rf{ghost-1} and \rf{2DTL-like} we can reexpress the 
action of
the $\pa$-pseudo-differential operators $\cM_s$ \rf{ghost-s} on EFs as
{\em ordinary $\bpa$-differential} operators, namely:
\begin{lemma}
For any generic eigenfunction $F$ of $\cL\,$, which does not appear within the
set $\lcurl \P_j \rcurl$ in \rf{ghost-s} (and whose ``ghost'' symmetry flows are
given by last eq.\rf{M-s-generic}) we have:
\be
\partder{}{\bt_s} \( F/\P_1\) = {\bar M}_s \( F/\P_1\)
\lab{ghost-s-F-P1} 
\ee
\be
{\bar M}_s \equiv 
\sum_{j=1}^s \biggl(\sum_{k=j}^s \frac{\P_{s-k+1}}{\P_1} \,
\frac{\cW_{j-1} (\Psi_k )}{\cW_j} \biggr) 
\(\bD - \bpa\ln \frac{\cW_{j-1}}{\cW_{j-2}} \) \cdots 
\(\bD - \bpa\ln \Psi_1\) \(\bD - \bpa\ln {1\o {\P_1}}\) 
- \partder{}{\bt_s} \ln \P_1  
\lab{bM-s}
\ee
where the $\bpa$-differential operators ${\bar M}_s$  satisfy the standard 
form of Zakharov-Shabat (ZS) ``zero-curvature'' equations w.r.t. the 
$\bt_s$-flows:
\be
\partder{}{\bt_s}{\bar M}_r - \partder{}{\bt_r} {\bar M}_s - 
\Sbr{{\bar M}_s}{{\bar M}_r} = 0
\lab{ZS-bM-s}
\ee
\label{lemma:M-s-bar}
\end{lemma}
Eq.\rf{ZS-bM-s} is a consequence of \rf{ZS-cM-s}.

According to ref.\ct{Takebe-92}, for any ZS system (as in 
\rf{ZS-bM-s}) there always exists a unique triangular
coordinate transformation in the space of evolution parameters such that
the (transformed) ZS differential operators acquire the standard KP form,
{\sl i.e.}, ${\bar M}_s = \(\bcL^s\)_{+}$ for some KP-type
$\bpa$-{\em pseudo-differential} operator $\bcL$. It turns out that
the ``ghost'' ZS operators \rf{bM-s} have already the right form:
\begin{proposition}
The $\bpa$-differential ZS operators ${\bar M}_s$ \rf{bM-s} can be expressed
as:
\br
{\bar M}_s = \(\bcL^s\)_{+} \quad ,\quad
\bcL 
= \bD + \sum_{k=1}^\infty b_k \,\(\bD + \bpa\ln \frac{W_{k+1}}{W_k} \)^{-1}
\cdots \(\bD + \bpa\ln \frac{W_2}{\P_1} \)^{-1}
\lab{bar-L}  \\
b_1 = -\bpa \(\P_2/\P_1\)     \quad ,\quad
b_k = -\bpa\(\P_{k+1}/\P_1\) + \ldots \quad {\rm for} \;\;
k=2,3,\ldots  
\lab{b-j-Phi}
\er
where the short-hand notations \rf{W-notation} are used.
\label{proposition:bar-L}
\end{proposition}

\mark
The pseudo-differential series of the original Lax operator $\cL$
\rf{lax-eq} can always be rearranged into a form similar to \rf{bar-L} :
\be
\cL = D + \sum_{k=1}^\infty a_k \,\( D -\pa\ln\frac{\cW_{k+1}}{\cW_k} \)^{-1}
\cdots \( D - \pa\ln \frac{\cW_2}{\Psi_1} \)^{-1}
\lab{L} 
\ee
with appropriate coefficients $a_k$. Expressions \rf{L},\rf{bar-L} are very
suggestive when discussing \DB (DB) orbits and connection to 2DTL to which
we now turn our attention.

Let us first introduce the following non-standard orbit of successive DB
transformations for the original KP system: 
\br
\cL (n+1) = T_1 (n) \cL (n) T_1^{-1} (n) \quad ,\quad
T_1 (n) = \P_1 D \P_1^{-1} \equiv \P_1^{(n)} D {\P_1^{(n)}}^{\, -1} 
\lab{DB-L} \\
\P_l^{n+1)} = \P_1^{(n)} \pa\(\frac{\P_{l+1}^{(n)}}{\P_1^{(n)}}\) 
\;\; ,\; l \geq 1 \; ;
\;\;\; \Psi_1^{(n+1)} = \frac{1}{\P_1^{(n)}} \;\; ,\;\;
\Psi_j^{(n+1)} = - \frac{1}{\P_1^{(n)}} \pai \(\P_1^{(n)} \Psi_{j-1}^{(n)}\) 
\; ,\; j \geq 2
\lab{DB-eigenf} \\
\cL (n-1) = {\hat T}_1 (n) \cL (n) {\hat T}_1^{-1} (n) \quad ,\quad
{\hat T}_1 (n) = \Psi_1 D \Psi_1^{-1} \equiv 
\Psi_1^{(n)} D {\Psi_1^{(n)}}^{\, -1} 
\lab{adjDB-L} \\
\P_1^{(n-1)} = \frac{1}{\Psi_1^{(n)}} \;\; , \;\;
\P_l^{(n)} = \frac{1}{\Psi_1^{(n)}} \pai \(\Psi_1^{(n)} \P_{l-1}^{(n)}\) 
\;\; ,\; l \geq 2   \; ;\;\;\; 
\Psi_j^{(n-1)} = - \Psi_1^{(n)} \pa \(\frac{\Psi_{j+1}^{(n)}}{\Psi_1^{(n)}}\)
\;\; , \; j \geq 1
\lab{adjDB-eigenf} 
\er
In what follows, the DB ``site'' index $(n)$ on (adj-)EFs will be skipped
for brevity whenever this would not lead to ambiguities.

\mark
Let us stress the non-canonical form of the (adjoint) DB transformations
\rf{DB-eigenf},\rf{adjDB-eigenf} on the ``ghost'' symmetry generating 
(adj-)EFs. For a generic eigenfunction $F$ the (adjoint) DB 
transformations read as usual \ct{oevela} :
\be
F^{(n+1)} = \P_1^{(n)} \pa \(\frac{F^{(n)}}{\P_1^{(n)}}\) 
= \(\pa - \pa\ln \P_1^{(n)}\) F^{(n)} 
\quad ,\quad
F^{(n-1)} = {1\o {\Psi_1^{(n)}}} \pai \(\Psi_1^{(n)} F^{(n)}\)
\lab{DB-generic}
\ee

We obtain the following important:
\begin{proposition}
``Ghost'' symmetries \rf{ghost-s} commute with DB transformations
\rf{DB-L}--\rf{adjDB-eigenf}, i.e., the ``ghost'' symmetry generators
\rf{ghost-s}
$\cM_s \equiv \cM_s (n) = \sum_{j=1}^s \P_{s-j+1}^{(n)} D^{-1} \Psi_j^{(n)}$
transform on the DB-orbit as:
\be
\cM_s (n) \quad \longrightarrow \quad
\cM_s (n\pm 1) =
\stackrel{(\wedge )}{T}\!\! (n) \cM_s (n) 
{\stackrel{(\wedge )}{T}}^{-1}\!\!\!\!\! (n)
+ \partder{}{\bt_s}\!\!\stackrel{(\wedge )}{T}\!\! (n) \,
{\stackrel{(\wedge )}{T}}^{-1}\!\!\!\!\! (n)
= \sum_{j=1}^s \P_{s-j+1}^{(n\pm 1)} D^{-1} \Psi_j^{(n\pm 1)}
\lab{DB-cM-s}
\ee
The ``ghost'' KP Lax operator \rf{bar-L} transforms, accordingly, as:
\be
\bcL(n+1) = \( {1\o {\P_1^{(n+1)}}} \bD^{-1} {\P_1^{(n+1)}}\)
\bcL(n) \( {1\o {\P_1^{(n+1)}}} \bD {\P_1^{(n+1)}}\)  
\lab{DB-bar-L} 
\ee
\label{proposition:DB-ghost}
\end{proposition}
We now are able to introduce double-KP hierarchy and its tau-functions.
We first construct an infinite set of (adj-)EFs
$\(\bP_j ,\bPsi_j\)_{j=1}^\infty$ for the ``ghost'' Lax
operator $\bcL$ \rf{bar-L} in terms of the initial set of (adj-)EFs
$\(\P_j ,\Psi_j\)_{j=1}^\infty$ of $\cL$ defining the 
``ghost'' symmetry flows \rf{ghost-s}--\rf{M-s-generic}.
Taking $F\! = \! const$ in \rf{ghost-s-F-P1} we find that
$\bP_1^{(n)} \equiv {1\o {\P_1^{(n)}}} = \Psi_1^{(n+1)}$
is an EF of $\bcL(n)$ for any ``site'' $n$ on the DB-orbit 
\rf{DB-L}--\rf{adjDB-eigenf}.
Therefore, taking into account \rf{DB-bar-L} we deduce that
$\bPsi_1^{(n-1)} \equiv {1\o {\bP_1^{(n)}}} = \P_1^{(n)}$
is an adj-EF of $\bcL(n-1)$ again for any DB ``site'' $n$.
The rest of the (adj-)EFs $\bP_j ,\bPsi_j$ for $\bcL$
($j \geq 2$) is constructed in such a way that their DB-orbit will have the
following form consistent with the DB-orbit of $\bcL$ \rf{DB-bar-L} :
\be
\bP_j^{(n-1)} = \bP_1^{(n)} \bpa
 \(\frac{\bP_{j+1}^{(n)}}{\bP_1^{(n)}}\) 
\;,\; j \geq 1 \; ;\;\;
\bPsi_1^{(n-1)} = {1\o \bP_1^{(n)}} \;\;, \;\;
\bPsi_l^{(n-1)} = - {1\o \bP_1^{(n)}} \bpai
\( \bP_1^{(n)} \bPsi_{l-1}^{(n)}\) \;, \; l \geq 2
\lab{bar-DB-eigenf} 
\ee
\be
\bP_1^{(n+1)} = {1\o \bPsi_1^{(n)}} \;\; ,\;\;
\bP_l^{(n+1)} = {1\o \bPsi_1^{(n)}} \bpai
\(\bPsi_1^{(n)} \bP_{l-1}^{(n)}\) \; , \; l \geq 2 ; \;\;
\bPsi_j^{(n-1)} = - \bPsi_1^{(n)} \bpa
\(\frac{\bPsi_{j+1}^{(n)}}{\bPsi_1^{(n)}}\) \;,\; j \geq 1
\lab{bar-adjDB-eigenf} 
\ee
\be
{\bar F}^{(n+1)} = \bP_1^{(n)} \bpa \(\frac{{\bar F}^{(n)}}{\bP_1^{(n)}}\) =
\(\bpa - \bpa\ln\bP_1^{(n)} \){\bar F}^{(n)}
\lab{bar-DB-F}
\ee
where ${\bar F}$ is a generic EF of $\bcL$.

The explicit form of $\lcurl \bP_j ,\bPsi_j \rcurl$ reads using notations
\rf{W-notation}--\rf{W-notation-1} :
\br
\bP_l^{(n)} &\equiv& \bP_l = 
(-1)^l {1\o {\P_1}} \bpai \P_1 \Psi_1 \bpai 
\frac{\cW_2}{\Psi_1^2} \bpai \frac{\cW_3 \Psi_1}{(\cW_2 )^2} \bpai
\cdots \bpai \frac{\cW_{l-1} \cW_{l-3}}{(\cW_{l-2})^2} \;\;\; ,\;\; l\geq 2
\lab{bcL-eigenf} \\
\bP_1^{(n)} \equiv \bP_1 \eq {1\o {\P_1}} \quad ;\quad
\bPsi_j^{(n)} \equiv \bPsi_j = (-1)^{j-1} \frac{W_2}{\P_1} \bpai
\frac{W_3 \P_1}{(W_2 )^2} \bpai \cdots \bpai \frac{W_{j+1} W_{j-1}}{(W_j )^2}
\;\;\; ,\;\; j \geq 1
\lab{bcL-adj-eigenf} 
\er

\mark
For later use let us write down 
the $k$-step iteration of DB transformations on $\P_1^{(n)}, \bP_1^{(n)}$:
\be
\P_1^{(n-k)} = (-1)^{k-1} \frac{\cW_{k-1}}{\cW_k} \quad ,\quad
\bP_1^{(n+k)} = {1\o {\P_1^{(n+k)}}} = \frac{W_{k-1}}{W_k}
\lab{DB-1-k}
\ee

Collecting the above results we obtain:
\begin{proposition}
Both Lax operators -- the initial $\cL$ \rf{L} and the ``ghost'' one  $\bcL$ 
\rf{bar-L}, define a 
double-KP integrable system:
\be
\partder{}{t_r} \cL = \Sbr{\(\cL^r\)_{+}}{\cL} \quad , \quad
\partder{}{\bt_s} \cL = \Sbr{\cM_s}{\cL} \quad , \quad
\partder{}{\bt_s} \bcL = \Sbr{\(\bcL^s\)_{+}}{\bcL}
\quad , \quad
\partder{}{t_r} \bcL = \Sbr{{\bar \cM}_r}{\bcL}
\lab{double-Lax-eqs}
\ee
where $\cM_s$ was introduced in \rf{ghost-s} and ${\bar \cM}_r$ is its ``dual''
counterpart defined in terms of the $\bcL$ (adj-)EFs
\rf{bcL-eigenf}--\rf{bcL-adj-eigenf} :
${\bar \cM}_r = \sum_{i=1}^r \bP_{r-i+1} \bD^{-1} \bPsi_i$ .
Accordingly, for generic EFs $F,\, {\bar F}$ of $\cL$ and 
$\bcL$, respectively, 
we have:
\br
\partder{}{t_r} F = \(\cL^r\)_{+} (F) \quad ,\quad 
\partder{}{\bt_s} F = \cM_s (F) \quad ,\quad
\partder{}{\bt_s} \( F/\P_1\) = \(\bcL^s\)_{+} \( F/\P_1\)  
\lab{F-eqs} \\
\partder{}{\bt_s} {\bar F} = \(\bcL^s\)_{+} \({\bar F}\) \quad ,\quad 
\partder{}{t_r} {\bar F} = {\bar \cM}_r \({\bar F}\) \quad ,\quad
\partder{}{t_r} \(\P_1 {\bar F}\) = \(\cL^r\)_{+} \(\P_1 {\bar F}\)
\lab{F-bar-eqs}
\er
\label{proposition:double-KP}
\end{proposition}

\begin{corollary}
According to Prop.\ref{proposition:double-KP} and \rf{ghost-s},
there exists a duality mapping between the two scalar KP subsystems of
\rf{double-Lax-eqs} defined by $\cL$ and $\bcL$, respectively, under
the exchange $(t) \leftrightarrow (\bt )$, $\P_j \leftrightarrow \bP_j$,
$\Psi_j \leftrightarrow \bPsi_j$ .
\label{corollary:duality}
\end{corollary}

There exists a simple relation between the tau-functions of $\cL$ and
$\bcL$. Namely, using \rf{M-s-eigenf} and \rf{ghost-1} in eq.\rf{bM-s}
for $s=2$ leads to: 
${\bar M}_2 \equiv \bcL^2_{+} = \bpa^2 - 2 \bpa \({\P_2}/{\P_1}\)$,
{\sl i.e.}, $\Res_{\bpa} \bcL = \bpa^2\ln {\bar \t} = - \bpa\({\P_2}/{\P_1}\)$ 
~which, upon comparing with the second eq.\rf{tau-P-bt-1}, implies for
${\bar \t}$ of $\bcL$ : $\bpa^2 \ln {\bar \t} = \bpa^2 \ln \(\P_1 \t\)$.
Applying duality (Corollary \ref{corollary:duality})
to the last equation and to first eq.\rf{tau-P-bt-1} we have also:
$\pa^2 \ln {\bar \t} = \pa^2 \ln \(\P_1 \t\)$ and
$\pa\bpa \ln {\bar \t} = \pa\bpa\ln \(\P_1 \t\)$.
The above relations can be generalized to the following:
\begin{proposition}  
The $\t$-function of $\bpa$-Lax operator $\bcL$ \rf{bar-L} is expressed in
terms of EFs and the $\t$-function of the original $\pa$-Lax operator $\cL$
\rf{L} as follows : 
\be
{\bar \t}(t,t^\pr ) = \P_1 (t,t^\pr ) \,\t (t,t^\pr ) \qquad ,\quad 
\frac{p_s \( -[\bpa]\) {\bar \t}}{{\bar \t}} = \frac{\P_{s+1}}{\P_1}
\lab{bar-tau}
\ee
\label{proposition:bar-tau}
\end{proposition}

Recalling expressions \rf{W-main} we derive from the second 
eq.\rf{bar-tau} a remarkably simple explicit
parametrization of the second ``ghost'' KP subsystem of \rf{double-Lax-eqs}
in terms of EFs of the first initial KP system:
\begin{corollary}
The dressing operator for $\bcL$ \rf{bar-L} has the following explicit form :
\be
\bcL = {\bar W} \bD {\bar W}^{-1} \quad ,\quad
{\bar W} = 1 + \sum_{s=1}^\infty \frac{\P_{s+1}}{\P_1} \bD^{-s}
\lab{W-bar}
\ee
\label{corollary:W-bar}
\end{corollary}
We now turn to construction of 2D Toda lattice hierarchy as a special DB-orbit 
of the double-KP system \rf{double-Lax-eqs}.
Recalling eqs.\rf{DB-1-k} allows to rewrite Lax operator expressions 
\rf{L},\rf{bar-L} in the form (reintroducing the DB ``site'' index) :
\br
\cL \equiv \cL (n) = D + \sum_{k=1}^\infty a_k (n) 
\( D - \pa\ln\P_1^{(n-k)}\)^{-1} \cdots \( D - \pa\ln\P_1^{(n-1)}\)^{-1}
\lab{cL-Toda} \\
{\bar \cL} \equiv {\bar \cL}(n) = \bD + \sum_{k=1}^\infty b_k (n) \,
\(\bD - \bpa\ln{\bar \P}_1^{(n+k)}\)^{-1} \cdots
\(\bD - \bpa\ln{\bar \P}_1^{(n+1)}\)^{-1}
\lab{bar-L-n}
\er
Now, taking into account \rf{DB-generic} and \rf{bar-DB-F},
one can represent the action of $\cL (n), \bcL (n)$  
\rf{cL-Toda}--\rf{bar-L-n} on generic EFs at any fixed ``site'' $n$ of the
DB-orbit as action of infinite Jacobi-type matrices $Q_{nm}$, ${\bar Q}_{nm}$ on
infinite column vectors 
$F^{(n)}$ and ${\bar F}^{(n)}$ ($k \geq 1$ below) : 
\be
\cL (n) \( F^{(n)}\) = Q_{nm} F^{(m)} \quad , \quad   
\bcL (n) \({\bar F}^{(n)}\) = 
\({1\o {\P_1^{(n)}}}{\bar Q}_{nm} \P_1^{(m)}\) {\bar F}^{(m)}
\lab{L-Q} 
\ee
\be
Q_{n,n+k} = \d_{k1} \quad ,\quad Q_{n,n-k} = a_k (n) \quad ,\quad
Q_{nn} = \pa\ln\P_1^{(n)}
\lab{Q-matrix} 
\ee
\be
{\bar Q}_{n,n-k} = \d_{k1} \frac{\P_1^{(n)}}{\P_1^{(n-1)}} = \P_1 \Psi_1 
\;\; ,\;\;
{\bar Q}_{n,n+k} = b_k (n) \frac{\P_1^{(n)}}{\P_1^{(n+k)}} =
b_k \frac{\P_1 W_{k-1}}{W_k}  \;\;,\;\;
{\bar Q}_{nn} = - \bpa\ln\P_1^{(n)}
\lab{bQ-matrix}
\ee
Using \rf{L-Q} the (pseudo-)differential Lax equations of the
double-KP hierarchy \rf{double-Lax-eqs}--\rf{F-bar-eqs}
for any fixed DB ``site'' $n$ can be equivalently represented as discrete
Lax equations for the infinite Jacobi-type matrices
\rf{Q-matrix}--\rf{bQ-matrix} :
\be
Q_{nm} \psi_m = \l \psi_n \quad ,\quad 
\partder{}{t_r} \psi_n = \(Q^r_{+}\)_{nm} \psi_m \quad , \;\;
\partder{}{\bt_s} \psi_n = - \({\bar Q}^s_{-}\)_{nm} \psi_m
\lab{2DTL-psi} 
\ee
\be
\partder{}{t_r} Q = \Sbr{\(Q^r\)_{+}}{Q}  \;\; , \;\;
\partder{}{\bt_s} Q = \Sbr{Q}{\({\bar Q}^s\)_{-}} \;\; , \;\;
\partder{}{\bt_s} {\bar Q} = \Sbr{{\bar Q}}{\({\bar Q}^s\)_{-}}
\;\; , \;\;
\partder{}{t_r} {\bar Q} = \Sbr{\(Q^r\)_{+}}{{\bar Q}}
\lab{2DTL-Lax}
\ee
where we took BA function as $F$, {\sl i.e.},
$F^{(n)} (t) = \psi_{BA}^{(n)} (t,\l ) \equiv \psi_n$ and
the subscripts $(\pm )$ indicate upper+diagonal/lower-triangular part 
of the corresponding matrices.
The above equations are the Lax equations for the 2DTL hierarchy
\ct{U-T} (see also \ct{BX}).

In this letter we announced our results on a new ``ghost'' symmetry
structure  
of the KP system giving rise to a duality between two related KP hierarchies
embedded into a {\em double-}KP system.   
Detailed exposition with complete proofs will appear elsewhere.
It will also address a variety of further
interesting issues: (a) Relation (embedding into) of the present double-KP
hierarchy \rf{double-Lax-eqs} to multi-component matrix KP hierarchies 
\ct{U-T}; (b) Generalization of the present construction with several infinite
sets of ``ghost'' symmetries; (c) Relation to random multi-matrix models 
\ct{BX}; (d) Supersymmetric generalization and obtaining a consistent 
supersymmetric 2DTL hierarchy.

\end{document}